\documentclass{ws-procs9x6-cpt16}
\begin{document}

\newcommand{\refeq}[1]{(\ref{#1})}
\def\etal {{\it et al.}}

\title{Astroparticles and Tests of Lorentz Invariance}

\author{J.S.\ D\'\i az}

\address{Institute for Theoretical Physics, Karlsruhe Institute of
Technology\\
76128 Karlsruhe, Germany}

\begin{abstract}
Searches for violations of Lorentz invariance using cosmic rays, gamma rays, and
astrophysical neutrinos and the prospects for future tests using cosmic-ray showers are presented.
\end{abstract}

\bodymatter

\section{Introduction}
The study of energetic particles bombarding Earth from distant astrophysical sources has led to the development of a new discipline: astroparticle physics.
The high energy and the long propagation distance of these astroparticles can serve as a sensitive tools to search for new physics, 
as minute unconventional effects can get enhanced by the energy and the path length.

\section{Cosmic rays and gamma rays}

A simple modification of quantum electrodynamics (QED) is obtained by incorporating a Lorentz-violating operator that preserves CPT, coordinate, and gauge invariance in the form\cite{SME}
\begin{equation}\label{L}
\mathcal{L} = -\frac{1}{4} F^{\mu\nu}F_{\mu\nu}
+ \overline{\psi}\big[\gamma^\mu(i\partial_\mu-eA_\mu)-m\big]\psi
-\frac{1}{4}(k_F)_{\mu\nu\rho\sigma} F^{\mu\nu}F^{\rho\sigma},
\end{equation}
where the first two terms correspond to conventional QED,
while the last term is a dimension-four operator for Lorentz violation in the Standard-Model Extension (SME).\cite{SME}
The nine independent components of the tensor $(k_F)_{\mu\nu\rho\sigma}$ that produce nonbirefringent effects have been constrained using the observation of high-energy cosmic rays.\cite{KlinkhamerRisse1,KlinkhamerRisse2}
In particular,
the isotropic limit is controlled by a single parameter $\kappa$,
whose upper limit has been constrained by the observation of high-energy cosmic rays,
whereas a lower limit has been obtained from energetic gamma rays.\cite{KS2008}

For $\kappa>0$,
an effective refractive index is produced,
which allows the production of Cherenkov radiation in vacuum.
This is an efficient energy-loss mechanism for electrically charged fermions above a threshold energy $E_{\mathrm{th}}$.
Cosmic-ray primaries would rapidly lose energy and fall below the threshold,
so that the emission of Cherenkov photons would rapidly stop.
Hence,
cosmic rays reaching Earth will always have energies below the threshold $E<E_{\mathrm{th}}$,
which allows constraining the positive range of values of $\kappa$.
The most stringent limit is $\kappa<6\times10^{-19}$ (98\% C.L.).\cite{KS2008}
Treating cosmic-ray primaries as point particles described in QED is a restrictive approximation because hadronic cosmic-ray primaries are composite particles.
In order to perform a more realistic description of proton primaries,
we considered the Cherenkov photon to be emitted by the charged constituents of the proton instead.\cite{Diaz:2015}
The calculation involves the determination of the power radiated by the quarks in the proton,
which is performed in the conventional way and then folded with the corresponding parton distribution functions.
A decrease in the power radiated is expected compared to the description of the full proton as a Dirac fermion because of the smaller electric charge of quarks and also due to the fact that charged partons carry only about half of the proton energy (the other half is carried by gluons).
Our results agree with these expected features;
nonetheless,
the total power radiated decreases only by one order of magnitude.
Given the astronomical distances traveled by cosmic rays,
the limits on $\kappa$ remain unaffected even in a realistic treatment of proton primaries.\cite{Diaz:2015}

For $\kappa<0$,
a photon becomes unstable and rapidly decays into an electron-positron pair.
This process corresponds to an efficient extinction mechanism for photons above a threshold energy $\omega_{\mathrm{th}}$ because astrophysical gamma rays with energies above the threshold would rapidly decay into lepton pairs.
Hence,
gamma rays reaching Earth will always have energies below the threshold $\omega<\omega_{\mathrm{th}}$,
which allows constraining the negative range of values of $\kappa$.
The most stringent limit is $\kappa>-9\times10^{-16}$ (98\% C.L.).\cite{KS2008}
Prospects for potential improvement on this lower limit could make use of cosmic rays by dedicated studies of the development of extensive air showers in the atmosphere,
whose maximum is very sensitive to the value of $\kappa$.\cite{Diaz:2016d}

\section{Astrophysical neutrinos}

Lorentz-violating neutrinos and antineutrinos in the SME are effectively described by a $6\times6$ hamiltonian of the form\cite{KM:2004,KM:2012}
\begin{equation}\label{Hnu}
H = |\pmb{p}|\begin{pmatrix} \mathbf{1} & 0 \\ 0 & \mathbf{1}
\end{pmatrix}
+ \frac{1}{2|\pmb{p}|}\begin{pmatrix} \mathbf{m}^2 & 0 \\ 0 & \mathbf{m}^{2*}
\end{pmatrix}
+ \frac{1}{|\pmb{p}|}\begin{pmatrix} \hat{\pmb{a}}_\mathrm{eff} -\hat{\pmb{c}}_\mathrm{eff}&& -\hat{\pmb{g}}_\mathrm{eff} + \hat{\pmb{H}}_\mathrm{eff}\\ -\hat{\pmb{g}}_\mathrm{eff}^\dag+\hat{\pmb{H}}_\mathrm{eff}^\dag && -\hat{\pmb{a}}_\mathrm{eff}^T-\hat{\pmb{c}}_\mathrm{eff}^T
\end{pmatrix},
\end{equation}
where each block is a $3\times3$ matrix in flavor space.
The first two terms appear in the Lorentz-invariant description of massive neutrinos of momentum $\pmb{p}$ characterized by a mass-squared matrix $\mathbf{m}^2$.
The last term includes all the modifications introduced by Lorentz violation.
Each matrix component depends in general on the neutrino momentum,
direction of propagation,
and the relevant coefficients controlling Lorentz violation.\cite{KM:2012}
In particular,
coefficients arising from a dimension-three operator lead to CPT violation that can mimic nonstandard interactions\cite{Diaz:2015b} and even the cosmic neutrino background.\cite{Diaz:2016a}

The effective hamiltonian \eqref{Hnu} has been used for the formulation of novel models describing global neutrino-oscillation data.\cite{SMEmodels} 
This hamiltonian also has served for implementing generic searches of the key signatures of Lorentz violation in experiments\cite{DataTables} using neutrino oscillations,\cite{KM:2004R,Diaz:2009,LV_LSND,LV_MINOS,LV_IceCube,LV_MiniBooNE:2012,LV_DC,LV_MiniBooNE:2013,Rebel:2013,Diaz:2013b,LV_SuperK} beta decay,\cite{Diaz:2013a,Diaz:2014d} and double beta decay.\cite{Diaz:2014b,EXO:2016}

Astrophysical neutrinos of very high energy have been observed by IceCube,\cite{Aartsen:2013}
which can be used to determine stringent limits on coefficients for Lorentz violation that modify the kinematics of neutrinos but are unobservable in oscillation experiments.
These oscillation-free coefficients can be constrained using an approach similar to the one presented in the previous section for cosmic rays and gamma rays.
For isotropic operators of arbitrary dimension $d$,
the relevant coefficients are denoted by ${\mathaccent'27 c}_\text{of}^{(d)}$.\cite{KM:2012}
For ${\mathaccent'27 c}_\text{of}^{(d)}<0$,
an effective refractive index is produced for neutrinos.
This allows the Cherenkov production of $Z$ bosons,
which rapidly decay into electron-positron pairs.
This is an efficient energy-loss mechanism for neutrinos above a threshold energy $E_{\mathrm{th}}$.
Astrophysical neutrinos would rapidly lose energy and fall below the threshold,
so that the Cherenkov emission would rapidly stop.
Hence,
high-energy astrophysical neutrinos reaching Earth will always have energies below the threshold $E<E_{\mathrm{th}}$,
which allows constraining the negative range of values of ${\mathaccent'27 c}_\text{of}^{(d)}$.\cite{Diaz:2014a,Diaz:2014c}
The observation of multiple events distributed in the sky allows also the study of anisotropic operators.\cite{Diaz:2014a}
Furthermore,
flavor-mixing operators could be studied by sensitive measurements of the flavor ratios of astrophysical neutrinos.\cite{Arguelles:2015} 

At low energies,
antineutrinos from the supernova SN1987A have been used to constrain dispersion effects produced by oscillation-free operators of dimension $d>4$.\cite{KM:2012}
Regarding the sensitive interferometric measurements of neutrino oscillations,
the absence of an antineutrino component in the flux of solar neutrinos can be used to determine the most stringent limits on Majorana couplings for CPT violation in the neutrino sector of the SME that would produce neutrino-antineutrino oscillations.\cite{Diaz:2016c}

\newpage
\section*{Acknowledgments}
This work was supported in part by the German Research Foundation (DFG) under Grant No. KL 1103/4-1.

\end{document}